\documentclass{jfm}

\usepackage{graphicx}
\usepackage{newtxtext}
\usepackage{newtxmath}
\usepackage{natbib}
\usepackage{hyperref}
\hypersetup{
    colorlinks = true,
    urlcolor   = blue,
    citecolor  = black,
}

\newcommand{\RomanNumeralCaps}[1]
\linenumbers
 
\newcommand{\Pra}{\mbox{\textit{Pr}}}  
\newcommand{\Ray}{\mbox{\textit{Ra}}}  
\newcommand{\Nus}{\mbox{\textit{Nu}}}  
\newcommand{\Ric}{\mbox{\textit{Ri}}}  

\title{Layering and vertical transport in sheared double diffusive convection in the diffusive regime}

\author{
  Yantao Yang\aff{1}\corresp{\email{yantao.yang@pku.edu.cn}},
  Roberto Verzicco\aff{2,3},
  Detlef Lohse\aff{2,4}
  \and 
  C. P. Caulfield\aff{5,6}\corresp{\email{cpc12@cam.ac.uk}}
}

\affiliation{
\aff{1}SKLTCS and Department of Mechanics and Engineering Science, BIC-ESAT,
       College of Engineering, and Institute of Ocean Research, 
       Peking University, Beijing 100871, China
\aff{2}Physics of Fluids Group, Department of Science and Technology, 
       Mesa+ Institute, Max-Planck Center Twente for Complex Fluid Dynamics, 
       and J. M. Burgers Center for Fluid Dynamics, University of
       Twente, 7500 AE Enschede, The Netherlands
\aff{3}Dipartimento di Ingegneria Industriale, University of Rome ``Tor Vergata'',
       Via del Politecnico 1, Roma 00133, Italy
\aff{4}Max Planck Institute for Dynamics and Self-Organisation, 37077 G\"{o}ttingen, Germany
\aff{5}BP Institute, University of Cambridge, Madingley Road, Cambridge CB3 0EZ, UK
\aff{6}Department of Applied Mathematics and Theoretical Physics, University of Cambridge,
       Wilberforce Road, Cambridge CB3 0WA, UK
}

\begin{document}

\maketitle

\begin{abstract}
A sequence of two and three-dimensional simulations is conducted for the double diffusive convection (DDC) flows in the diffusive regime subjected to an imposed shear. The flow is confined between two horizontal plates which are maintained at different constant temperature, salinity, and different velocity, thus setting up a shear across the flow. The lower plate is fixed at higher temperature and salinity, while the overall (unperturbed) density gradient is statically stable. For a wide range of control parameters, and for sufficiently strong perturbation of the conductive initial state, we find that staircase-like structures spontaneously develop, with relatively well-mixed layers separated by sharp interfaces of enhanced scalar gradient. Such staircases appear to be robust even in the presence of strong shear over very long times, although we typically observe early time coarsening of the number of observed layers. For the same set of control parameters, different asymptotic layered states, with markedly different vertical scalar fluxes, can arise for different initial perturbation structures. The imposed shear does significantly spatio-temporally modify the vertical transport of the various scalars. The flux ratio $\gamma^*$ (i.e., the ratio between the density fluxes due to the total (convective and diffusive) salt flux and the total heat flux) is found, at steady state, to be essentially equal to the square root of the ratio of the salt diffusivity to the thermal diffusivity, consistently with the physical model originally proposed by \cite{Linden1978} and the variational arguments presented by \cite{Stern1982} for unsheared double diffusive convection.
\end{abstract}


\section{Introduction}

Multi-component fluids, where the density changes due to variations in the concentrations of two scalars with differing diffusivities are prone in many circumstances to  rich dynamical behaviour, commonly referred to as ``double-diffusive convection'' (DDC). DDC, as originally described by \cite{Stern1960}, occurs in a wide range of geophysical, industrial and astrophysical contexts, as reviewed for example by~\cite{Turner1985,Schmitt1994,Radko2013,Garaud2018}. A particular application of profound interest arises in certain parts of the world's oceans, where both heat and salt concentration (i.e. salinity) lead to variations in fluid density. As the diffusivity of heat is $O(100)$ times the diffusivity of salt, seawater can be prone to DDC, which is usually divided into two broad classes. `Fingering' DDC occurs when relatively salty and warm fluid overlies relatively fresh and cold fluid, while conversely `diffusive' DDC occurs when relatively fresh and cold fluid overlies relatively salty and warm fluid. Interestingly, both classes can lead to the formation of double-diffusive `staircases', where relatively deep and homogeneous `layers' are separated by relatively thin `interfaces', exhibiting strong scalar gradients. 

Both classes of DDC, and the associated `staircases'  play a key role in vertical mixing-induced transport in the ocean. Such vertical transport is both a key driver of the global climate system \citep{Ferrari2009} and also an outstanding area of uncertainty in the modelling of our changing climate. Clearly, staircase structures in scalar fields will significantly affect vertical convective and diffusive scalar transport compared to flows with uniform gradient. Therefore, understanding how such staircases are `born' and `survive' in the presence of larger-scale flows is of great practical interest. Broadly speaking, each class of DDC can be associated with different regions of the world, where differing physical processes lead to the required opposing gradients of heat and salinity. Specifically, fingering DDC typically arises in the tropics, where thermally-driven evaporation leads to the required relatively salty and warm fluid overlying  fresher and cooler water. \cite{Schmitt2005} demonstrated through observational measurement that fingering-induced staircases are associated with vigorous turbulent mixing and hence crucially significant for vertical scalar transport in the Caribbean, with apparently a larger net contribution than the contribution due to internal wave breaking.

Conversely, diffusive DDC is common at high latitudes, where the inevitable interaction with ice and meltwater leads to  the required conditions for  diffusive DDC, which has a major influence both on the climate of polar regions, and larger-scale oceanic circulations, as discussed by \cite{Turner2010,Shaw2014,Bebieva2017,Bebieva2019a,Bebieva2019b}. The climate, particularly in the Arctic, is rapidly changing (as reviewed by \cite{Timmermans2020}) while diffusive staircases are known, at least at present, to play a dominant role in vertical mixing, and thus act as a critical chokepoint for heat transport upwards of relatively warm Atlantic waters.

Therefore, understanding the dynamical properties of how such diffusive DDC staircases can form, and (at least equally importantly) remain robust is a topic of not only fluid dynamical interest but also great climatological importance, and so we focus on such staircases here. Diffusive staircases have been observed to develop spontaneously, even in numerical simulations restricted to two dimensions both with and without shear \citep{Noguchi2010,Zaussinger2018}, but their actual formation mechanism has up to now remained somewhat mysterious, as discussed by~\cite{Kelley2003,Radko2013}. More recently, highly plausible candidate mechanisms for both the  formation and apparent survival  of such staircases have been proposed, exploiting  a flow characteristic that could reasonably be assumed to be generically present: namely shear \citep{Radko2016,Radko2019a}. Indeed, \cite{Radko2016,Radko2019a} beautifully demonstrated that an appropriate combination of vertical shear and multi-component diffusion, individually linearly stable, is prone to a linear `thermohaline-shear' instability, that leads, at finite amplitude, to a staircase-like structure of relatively well-mixed and deep layers separated by relatively thin interfaces with strong scalar gradients. 

Furthermore, \cite{Brown2019} have shown that such an instability leading to layering can continue to arise when the shear is time-dependent, as might be expected in the presence of sufficiently weak internal waves. This study was followed up by direct numerical simulation of a single interface with an imposed linear shear by \cite{Brown2021} within a vertically periodic domain, demonstrating rich and complex dynamics in the vicinity of the energised interface. Such interface was observed to `survive' in the range of parameters that they considered. Using a reduced modelling approach, \cite{Shibley2019} demonstrated that diffusive staircases also appeared to survive in the presence of sufficiently weak ambient turbulence, plausibly again associated with some combination of shear-driven instabilities and breaking internal waves. An alternative formation mechanism relying on horizontal interleaving in the presence of non-trivial eddy diffusivities and viscosities has been proposed by \cite{Bebieva2017}, once again pointing to the important physical roles played by shear and turbulence in the staircase dynamics. 

However, it is still unclear how such layers evolve in the presence of both sufficiently vigorous turbulence and significant shear, as might be expected to occur frequently in the world's oceans. The absence of widespread consideration of flows with significant shear is particularly concerning, as there is generically at least some velocity shear below sea ice, where the presence of double diffusive staircases has been observed to suppress  vertical heat fluxes significantly, compared to an assumed purely (and in some sense homogeneous) turbulent transport (see for example \cite{Bebieva2019b}).

Of course, modelling can be challenging when the flow is inherently, and profoundly nonlinear. Even if the class of  flow considered is highly idealised, especially compared to actual oceanographic situations, there are still major technical obstacles which need to be overcome  for the analysis of the flow dynamics. Analytical progress is likely to be very challenging when the flow is highly spatio-temporally variable, to put it mildly. Numerical simulation of sheared diffusive DDC prone to layering is also exceptionally difficult for  two, inter-related,  reasons. First, simulation of turbulent flows with a `staircase' structure in a scalar field must accurately represent both `overturning' of relatively `weak' interfaces and `scouring' of relatively `strong' interfaces by impinging turbulent eddies, using the classification originally proposed by \cite{Woods2010},  building upon the classical insights of \cite{Kato1969}. Indeed, \cite{Brown2021} clearly observed `scouring' dynamics in the vicinity of the sheared interface, which they interpreted as evidence of the so-called `Holmboe wave instability' (HWI) \citep{Holmboe1962}. Such a HWI generically occurs in  inflectional shear flows across `sharp' density interfaces, and, at sufficiently high flow Reynolds number, can lead to significant turbulent transport across such interfaces, which still survive for significant periods of time without being fully eroded \citep{Salehipour2016}.

Crucially, as demonstrated by \cite{Taylor2017}, a particular structure of the effective diffusivity near the interface must be maintained so that it is not eroded, and so it is critical that numerical simulation captures the interfacial dynamics accurately, specifically not introducing significant spurious diffusivity which might erroneously `smooth out' interfacial structure. Essentially, this may be thought of as a requirement to capture the inherent `multi-scale' nature of such flows, where sharp interfaces need to be captured accurately while also relatively deep and horizontally extended layers must also be simulated. Evidence is building that such layer-interface staircase structures arise in many density-stratified situations (see for example \cite{Caulfield2021} for a review) and various insights have been gained into the necessity for sufficiently large dynamic range to exist between vertical scales significantly affected by stratification and viscosity \citep{Portwood2019}, even when there is a single diffusive scalar with scalar diffusivity of the same order as kinematic viscosity.

However, the second reason that  DDC is so challenging to simulate (as discussed for example by \cite{Brown2021})  is that there is inevitably a further hierarchy of scales which must be captured, due to the inherent requirement of differing diffusivities for the two scalar fields, i.e. heat and salinity in the oceans. Capturing the dynamics of the scalar field with smaller diffusivity (e.g. `salt') requires an even wider range of scales to be captured in a layered and sheared flow, which is exceptionally expensive computationally. Here, we address this challenge  by using a bespoke code~\citep{multigrid2015,ddcjfm2016} with `nested' grids, so that the velocity and `temperature'  fields are solved on a base mesh within which a much refined mesh is nested for the low-diffusivity salinity field. Using this code, we are now in the position to consider the long-time dynamics of diffusive DDC flows with three key characteristics:  turbulent convection, or at least vigorous disorder; susceptibility, at least in principle, to staircase structure in temperature, salinity and overall density; and large-scale imposed, and hence `forcing', shear.

As noted above, understanding vertical transport of scalars in oceanic situations prone to diffusive DDC is vital to larger-scale climate modelling, and the presence (or absence) of staircase structure is known to modulate this transport significantly. Therefore, the study presented here has two interlocking aims: 

First, we wish to understand the circumstances under which staircases can arise, and crucially survive, in diffusive DDC, subject to an imposed shear in an idealised flow geometry. As we shall see, rich dynamics of layer appearance, merger and coarsening are possible. Moreover, we demonstrate the co-existence of multiple, at least meta-stable, layered states for different initial conditions in flows with the same global control parameters. We are focussed on inherently nonlinear and turbulent flows. Therefore, we are not focussed on constructing a description in terms of linear instability or potentially a hierarchy of instabilities, but rather in terms of the robustness of basins of attraction of nonlinear system states.

Second, we wish to understand and quantify various key properties of the vertical scalar fluxes: of heat; of salt; and of mass, and the extent to which classical diffusive DDC behaviour is modified (or not) by the presence of mean shear. Since we have access to the entire numerical data, we are able to consider such properties in unprecedented spatio-temporal detail. Naturally, we can consider not only typical averaged values of appropriately defined Nusselt numbers (i.e. the enhancement of flux due to convection over purely diffusive values), but crucially their variation in space and time as the staircase structure also evolves. We also are able to investigate how `efficient' the mixing is, in the sense of the properties of the turbulent flux coefficient $\Gamma$, the ratio between the vertical density flux and the kinetic energy dissipation rate. This quantity was originally defined by \cite{Osborn1980} in single-scalar stratified flows. Of course, since we are considering sheared diffusive DDC, we expect the vertical density flux actually to be negative, as generically, diffusive DDC naturally tends on average to {\it reduce} the potential energy of the system, through the vertical upward flux of heat  outweighing the vertical downward flux of salt \citep{Radko2013}. Nevertheless, it is clearly of interest to investigate how this phenomenon is affected by shear, the extent to which spatio-temporal variability is significant, and also whether it is possible to parameterise the various scaled fluxes in terms of different non-dimensional parameters of the flow.

To achieve these two aims, the rest of the paper is organised as follows. In \S~\ref{sec:method}, we describe the flow geometry, the governing equations, and the numerical details. In \S \ref{sec:morph} and \ref{sec:transp} we then present our results about the physical properties of the observed layered states and the associated vertical transport. In \S \ref{sec:multi} we discuss the interesting observation that different layered states can be stable over long times within flows with the same global control parameters and the impact of multiple states on the global fluxes. Finally, in \S \ref{sec:conc}, we draw some brief conclusions.

\section{Governing equations and numerical methods}\label{sec:method}

We consider a fluid layer bounded by two parallel plates which are perpendicular to gravity and separated by a height $H$. We employ the Oberbeck-Boussinesq approximation, namely the fluid density depends linearly on both temperature and salinity as $\rho(\theta, s)=\rho_0[1 - \beta_\theta \theta + \beta_s s]$. Here $\rho_0$ is a reference value for density, $\beta_\zeta$ with $\zeta=\theta$ and $s$ are the positive expansion coefficients of two scalar components.
As in the motivating oceanographic application described in the introduction, these two scalar components can be thought of as heat and salt. The associated temperature $\theta=T-T_0$ and salinity $s=S-S_0$ are also relative to their respective reference values. We use the conventional geophysical coordinate system, so that gravity is directed in the (negative) $z-$direction, and $x$ and $y$ are the two horizontal directions parallel to the plates. For the diffusive regime of DDC, the lower plate is maintained at both higher temperature and higher salinity relative to the upper plate, and so the conductive temperature gradient is destabilising, while the conductive salinity gradient is stabilising. Fixed differences are sustained across the fluid layer in temperature $\Delta_\theta=\theta(0)-\theta(H)$ and in salinity $\Delta_s=s(0)-s(H)$, respectively. Therefore, a fixed density difference is also maintained across the layer $\Delta_\rho=\rho(0)-\rho(H)=\rho_0[\beta_s \Delta_s - \beta_\theta \Delta_\theta]$.

To introduce the external shearing, we decompose velocity into $\mathbf{u}^* = \mathbf{u} + \mathbf{U}_S = \mathbf{u} + (U_b/H) \left(z-H/2\right)\mathbf{e}_y$, where $\mathbf{e}_y$ is the unit vector in the (streamwise) $y$-direction. $U_b/H$ is the constant overall shear rate, of the background (constant and linear) velocity $\mathbf{U}_S$, while $\mathbf{u}$ is the perturbation velocity. The streamwise velocity at the top and bottom plates is fixed at the values of the background velocity $\pm U_b/2$, respectively.
The governing equations are
\begin{subequations}\label{eq:sddc}
\begin{eqnarray}
  \partial_t u_i + u_j \partial_j u_i + U_{Sj}\partial_ju_i + u_j\partial_j U_{Si} &=& 
       - \partial_i p + \nu \partial_j^2 u_i 
       + g \delta_{i3} (\beta_\theta \theta - \beta_s s), \label{eq:momem}  \\
  \partial_t \theta + u_j \partial_j \theta + U_{Sj}\partial_j \theta &=& 
       \kappa_\theta \partial_j^2 \theta,   \label{eq:temper}  \\
  \partial_t s + u_j \partial_j s + U_{Sj}\partial_j s &=&
       \kappa_s \partial_j^2 s, \label{eq:concen}
\end{eqnarray}
\end{subequations}
in which $u_i$ with $i=x,y,z$ are the three components of the perturbation velocity,  $p$ is pressure, $\nu$ is kinematic viscosity, $g$ is the gravitational acceleration, and $\kappa_\zeta$ are the molecular diffusivities of the two scalars.  The background velocity therefore exerts  an inertial body force on the flow. The perturbation velocity field also satisfies the continuity equation $\nabla\cdot\mathbf{u}=0$. Next to the strength of the shear force (expressed through the (inverse) Richardson number below in~\eqref{eq:ric}), the control parameters are the two Prandtl and the two Rayleigh numbers,
\begin{equation}
  \Pra_\zeta = \frac{\nu}{\kappa_\zeta}, \quad \quad 
  \Ray_\zeta = \frac{g \beta_\zeta \Delta_\zeta H^3}{\kappa_\zeta \nu}.\label{eq:prra}
\end{equation}
It must always be remembered that the background salinity jump induces a stabilizing buoyancy force.  The relative strength of this stabilizing buoyancy force induced by the salinity difference to the driving force of temperature difference can be expressed by the density ratio
\begin{equation}\label{eq:lambda}
  \Lambda = \frac{\beta_s \Delta_s}{\beta_\theta \Delta_\theta} 
  = \frac{\Pra_\theta \Ray_s}{\Pra_s \Ray_\theta}.
\end{equation}
The relative magnitude of the external buoyancy to the external shearing is represented by an appropriate (bulk) Richardson number, defined as
\begin{equation}\label{eq:ric}
  \Ric_b = \frac{g \Delta_\rho H}{\rho_0 U_b^2}=\frac{g \beta_\theta \Delta_\theta H}{U_b^2}
  \left ( \Lambda- 1 \right ) ,
\end{equation}
demonstrating that overall static stability with $\Ric_b \geq  0$ requires $\Lambda \geq 1$. 

We numerically solve the governing equations (\ref{eq:sddc}) for $\mathbf{u}$, $\theta$ and $s$ by using our efficient code~\citep{multigrid2015,ddcjfm2016}. A special feature of the code is that the momentum and temperature are solved on a base mesh while a significantly more refined mesh is used for the salinity field due to its significantly smaller diffusivity. The physical quantities are non-dimensionalized by $H$, the free-fall velocity associated with the statically unstable background temperature difference $\sqrt{g \beta_\theta \Delta_\theta H}$, and the two scalar differences $\Delta_\zeta$. Henceforth, all variables are non-dimensional, unless noted otherwise. Periodic boundary conditions are applied in the horizontal directions. At the two plates both the temperature and salinity are kept constant at their boundary values, and the perturbation velocity $\mathbf{u}$ obeys free-slip conditions in the horizontal, with no penetration $u_z(0) = u_z(1) =0$ in the vertical direction. In this study we set the two Prandtl numbers as $\Pra_\theta=10$ and $\Pra_s=1000$, which are largely consistent with those appropriate for seawater~\citep{Radko2016}. This means the diffusivity ratio $\tau$, is
\begin{equation}
  \tau = \kappa_s \left/ \kappa_\theta \right. = 0.01. \label{eq:taudef}
\end{equation}
To limit the control parameter space, we fix the density ratio at $\Lambda=2$, which is relevant to the high-latitude oceans~\citep{Timmermans2008,Turner2010,Shibley2017}. Meanwhile, we fix the bulk Richardson number at $\Ric_b=1$, which exceeds the threshold value $1/4$ for dynamical instability. To quantitatively describe the influences of shear strength, it is highly desired to simulate a series of different $\Ric_b$. Nevertheless, in the present study we focus on the rich dynamics of the layering and its impact on the vertical transport. Systematic study investigating $\Ric_b$-dependence is left for future work.

Details of the simulations are summarized in table~\ref{tab:num}. The mesh size is chosen to be smaller than the Kolmogorov scale for the velocity field and the Batchelor scale for the scalar field, thus all the relevant physical scales are adequately resolved. In total seven cases were conducted with various $Ra_\theta$. For the two smallest values of $Ra_\theta$, simulations were run in three dimensions (3D). While for the higher values of $Ra_\theta$, the simulations were restricted to two dimensions (2D) since the 3D simulations require too much computing resource. Moreover, for $Ra_\theta=10^6$ we have run both 3D and 2D simulations, and for different streamwise extent $L_y$, as listed as cases 2-4 in the table. The three cases exhibit similar flow evolution, including in terms of the various fluxes, and so we believe that 2D simulations can capture the essential dynamics of the system, at least within this flow geometry (c.f. \cite{Brown2021}). 
\begin{table}
  \begin{center}
  \def~{\hphantom{0}}
  \setlength{\tabcolsep}{8pt}
  \begin{tabular}{ccccccccc}
  3D Cases & $\Ray_\theta$ & $\Ray_s$ & $(L_x,~L_y)$ 
   &  $N_x,~N_y,~N_z$ & $ m_x(m_y),~m_z$ & $k$ & $\delta$ \\[3pt]
    1  & $10^5$  & $2\times10^7$ & $(2,4)$ & $240,~360,~240$ & $4,~3$ & 4 & 0.1 \\
    2  & $10^6$  & $2\times10^8$ & $(1,2)$ & $240,~360,~360$ & $4,~3$ & 6 & 0.05 \\[10pt]
    
  2D Cases & $\Ray_\theta$ & $\Ray_s$ & $L_y$  
   &  $N_y,~N_z$ & $m_y,~m_z$ & $k$ & $\delta$ \\[3pt]
    3  & $10^6$  & $2\times10^8$    & $2$ & $384,~384$   & $4,~3$ & 6 & 0.05 \\
    4  & $10^6$  & $2\times10^8$    & $4$ & $768,~384$   & $4,~3$ & 6 & 0.05 \\
    5  & $10^7$  & $2\times10^9$    & $4$ & $1440,~720$  & $4,~3$ & 8 & 0.025  \\
    6  & $10^8$  & $2\times10^{10}$ & $2$ & $1536,~1152$ & $4,~3$ & 16 & 0.01  \\
    7  & $10^7$  & $2\times10^9$    & $4$ & $1440,~720$  & $4,~3$ & 8 & 0.025  \\
  \end{tabular}
  \caption{Parameters and numerical details of simulated cases. Columns from left to right: case number, Rayleigh numbers of temperature and salinity, aspect ratios in horizontal directions, the grid size of the base mesh, the refinement factor for the refined mesh, and the parameters $k$ and $\delta$ for the initial perturbations used in (\ref{eq:init}) for cases 1-6.
Case 7, (with the same bulk parameters as case 5) has initial condition given by (\ref{eq:case5n}). For all cases, the density ratio is fixed at $\Lambda=2$ and the Richardson number at $Ri_b=1$. }
  \label{tab:num}
  \end{center}
\end{table}

To investigate the relevance of the simulated parameters to the ocean environment, we calculate the corresponding dimensional quantities for cases 5 and 6 listed in Table~\ref{tab:num}. We choose $\beta_\theta=6.3 \times 10^{-5} K^{-1}$, $\beta_s=7.8\times10^{-4} psu^{-1}$, $\kappa_\theta=1.4\times10^{-7} m^2/s$, $g=9.8m/s^2$, and a total temperature difference of $\Delta_\theta=0.1 K$, respectively. Then for the case 5 with $\Ray_\theta=10^7$ the height of the fluid layer is $H\approx0.3m$ and the total simulation time is about 5 days. For the case 6 with $\Ray_\theta=10^8$, the total height is about $H\approx0.7m$, and the simulation time is about two weeks.

Before proceeding to discuss the results, we first comment on the choice of the boundary condition and the initial conditions that appear to be required to trigger layering in our system. Unlike fully periodic domains, for the current vertically bounded model it is straightforward to impose a background shear with uniform strength. Due to the no penetration condition the scalar transport within a thin layer adjacent to the two boundaries is dominated by the conductive part, and the global fluxes are inevitably affected by the boundaries. However, since it is the total scalar difference which is fixed across the whole domain, the final global fluxes are the combined results of both the bulk dynamics and the conductive regions close to the boundary. In the following we will focus on the bulk region and investigate the behaviours of horizontally averaged local quantities relatively far from the boundaries. The global fluxes will mainly be used as quantitative indicators of the evolution and transition of the global flow morphology.

For a fully periodic domain, the so-called gamma-instability can explain the emergence of staircase and layering for fingering DDC~\citep{Traxler2011,Stellmach2011}. While for the diffusive regime, \cite{Radko2016} elegantly proved that linear instability occurs when the fluid layer experiences linear background scalar gradients and a vertical shearing with a sinusoidal velocity profile, even in a situation when the flow is stable to both shear-driven and convectively-driven instabilities. In our configuration, initially both scalars have vertically linear distributions, and the shearing strength is uniform within the fluid layer.  Due to the presence of the top and bottom boundaries, the linear instability mechanisms associated with the fully periodic domain are unlikely to be applicable to the current model. Indeed, linear instability analysis is conducted for $\Lambda=2$ and $Ri_b=1$ by using the standard normal-mode method. The results indicate that the current system is linearly stable, i.e. is stable to infinitesimal perturbations. However, as described in the following sections, numerical simulations reveal that, if the initial perturbation is sufficiently large in amplitude, nonlinear processes lead to escape from the basin of attraction of the linear `conductive' base flow state, with non-trivial motion and indeed layering being triggered in the flow. 

Specifically, the initial scalar distributions are given as
\begin{equation}\label{eq:initall}
  \theta_{\rm ini} = (1-z) - \theta_{\rm pert}, \quad
  s_{\rm ini} = (1-z) - s_{\rm pert}.
\end{equation}
For cases 1-6, the perturbation parts are
\begin{equation}\label{eq:init}
  \theta_{\rm pert} = \delta \sin(2 \pi k z), \quad
  s_{\rm pert} = \delta \sin(2 \pi k z), \quad \mbox{for} \quad 0\le z\le 1.
\end{equation}
The values of $\delta$ and $k$ for each case are given in table~\ref{tab:num}. Numerical tests suggest that, to start non-trivial flow and in particular layering, $\delta$ has to be sufficiently large so that a locally statically unstable stratification arises. As noted in the introduction, it is not appropriate to view the response of the system to such a strong perturbation as the growth of a linear instability, but rather as an identification of basin(s) of attraction of robust nonlinear states of the entire system. Taking this viewpoint further, we test whether there is a unique non-trivial attractor for each set of global flow control parameters. For case 7 we choose exactly the same bulk parameters as case 5 but with a different initial condition. In this new case only a local density inversion is prescribed in the middle of the domain, namely, similar to \eqref{eq:init}, but now with only one wave pattern centred around $z=1/2$, i.e.
\begin{equation}\label{eq:case5n}
  \theta_{\rm pert} = \delta \sin(2 \pi k z), \quad
  s_{\rm pert} = \delta \sin(2 \pi k z), \quad \mbox{for} \quad |z-1/2| \le 1/(2k).
\end{equation}
where $\delta$ and $k$ are given in table \ref{tab:num}, and importantly, are the same as for case 5.

We will present our results in three different ways. We first consider the flow morphology and its evolution (\S \ref{sec:morph}). We then focus on vertical transport in the layered state which develops in case 5, at the temperature Rayleigh number $Ra_\theta=10^7$  (\S \ref{sec:transp}). Finally, in \S \ref{sec:multi} we demonstrate that multiple asymptotic states exist for the same global flow control parameters, by considering in detail the behaviour of case 7. As said above, case 7 has the same parameters as case 5, but different initial conditions, with the perturbation localised at the horizontal midplane of the flow. The qualitatively different asymptotic behaviour of these two cases  shows by construction that the flow evolution can be very sensitive to the particular characteristics of the initial conditions.

\section{Flow morphology of different cases}\label{sec:morph}

\begin{figure}
  \centerline{\includegraphics[width=\textwidth]{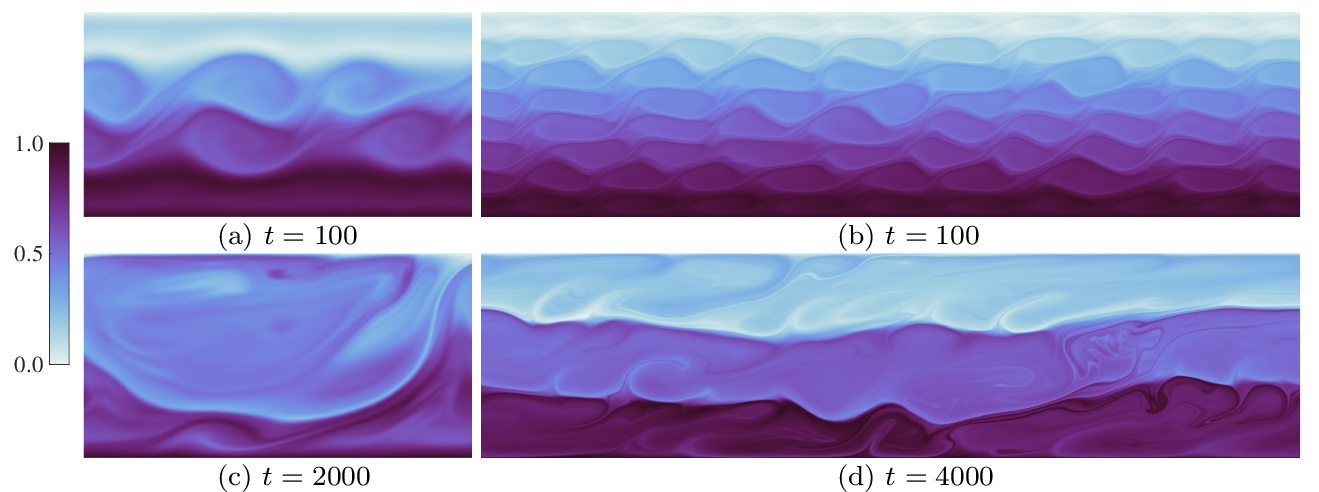}}
  \caption{Density fields at different stages of (a, c) case 1 with $Ra_\theta=10^5$ and (b, d) case 4 with $Ra_\theta=10^7$. Panels a and b show the flow fields at the early stage when the initial perturbations grow into layers of vortices. Panels c and d show those at the final stage with distinct layers separated by sharp interfaces with high density gradient. Three-dimensional data is from the $y-z$ plane at the $x-$midpoint of the computational domain. In both cases, the emergence of the layers and subsequently their coarsening with time is clearly seen. All panels share the same color map as shown at the left.}
\label{fig:density}
\end{figure}

In figure~\ref{fig:density} we show vertical cross-sections of density distributions for two cases. With the non-dimensionalisation described in the previous section, the density varies from one at the lower boundary $z=0$ to zero at the upper boundary $z=1$. The initially locally unstable stratification generates  convective flows, which in turn develop into spanwise vortices due to the vertical shearing, somewhat reminiscent of Kelvin-Helmholtz (KH) billows. These vortices in turn  evolve in a way that leads to  layering in the scalar field. For the 3D case 1, with the smallest considered $Ra_\theta$, these KH vortices occupy the whole channel. For the cases with higher $Ra_\theta$, the vertical extent of the vortices decreases. Multiple layers emerge, separated by relatively sharp interfaces, as shown for example in the lower right panel of figure~\ref{fig:density}. The sharp interfaces refer to the relatively thin regions of substantially enhanced scalar gradients. Clearly, two sharp interfaces exist for the 2D case 4 with $Ra_\theta=10^7$. They oscillate strongly in space and interact with the convection flow in the relatively well-mixed layers, but the interfaces remain robust. This behaviour can be clearly seen in the supplementary movies, showing for all cases the time evolution of the two scalars independently and in combination as the density. There are some structures which are reminiscent of the classical finite amplitude `cusped waves', the finite amplitude manifestation of the Holmboe wave instability, see for example \citet{Salehipour2016}. Similar phenomena are also noted by
\cite{Brown2021} in sheared diffusive DDC. Though, it is challenging to identify these structures categorically as Holmboe wave instability, as any impinging vortical structures are likely to induce such scouring motions in the vicinity of a sharp interface. 

Moreover, figure~\ref{fig:density} illustrates why very fine grids are needed for the current study. The interfaces with relatively high density gradient have very small vertical extent, while the convective plumes emerging from these interfaces are also very narrow, requiring a very high resolution. 

\begin{figure}
  \centerline{\includegraphics[width=\textwidth]{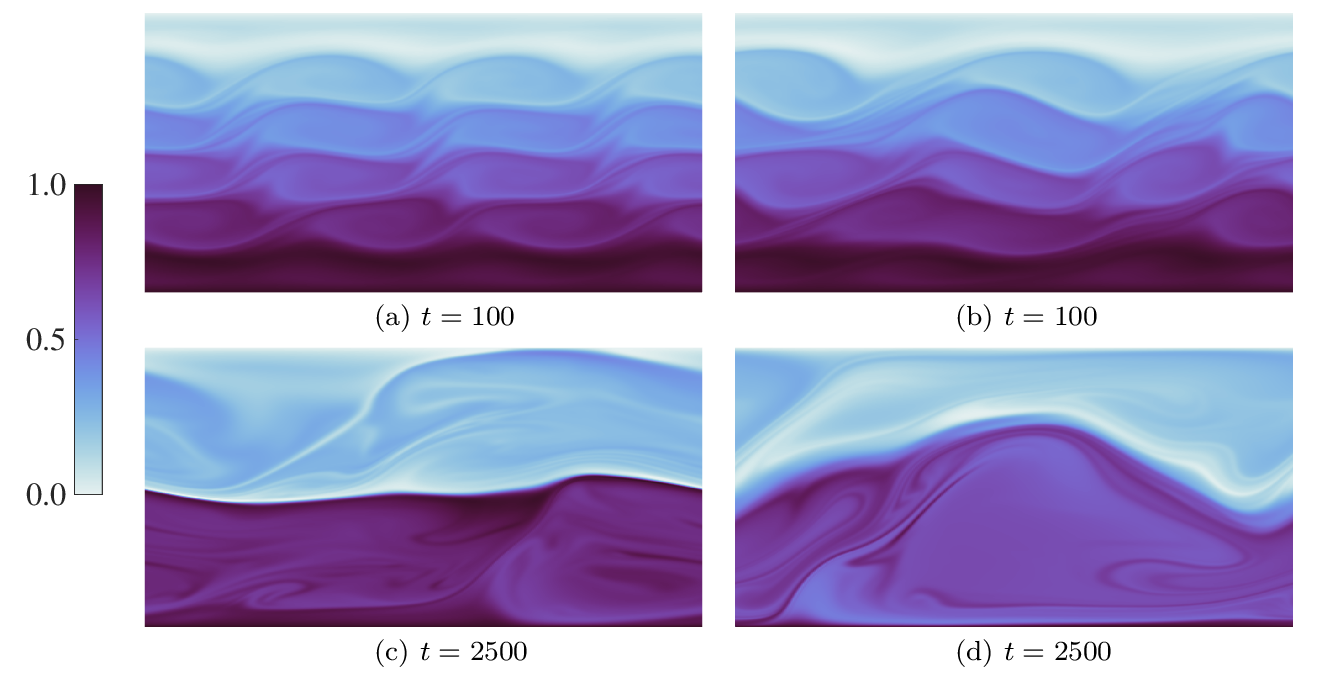}}
  \caption{Density fields from the 3D simulation of case 2 (a, c) and the 2D simulation of case 3 (b, d). The two cases have the same $Ra=10^6$. (a, b) are for time $t=100$ and (c, d) for $t=2500$. Three-dimensional data is from the $y-z$ plane at the $x-$ midpoint of the computational domain. All panels share the same color map as shown at the left.}
\label{fig:comp2d3d}
\end{figure}

To verify that 2D simulations exhibit similar dynamics to 3D simulations, we compare the 2D case 3 to the 3D case 2. Both cases have the same control parameters. Flow fields are compared in figure~\ref{fig:comp2d3d} at two stages: the  relatively early shear-induced KH vortices stage at $t=100$; and the later layering stage at $t=2500$. The flow morphology is largely similar in two and three dimensions. For both simulations a single sharp interface remains during the late layering stage. Though the oscillation of interfaces is stronger in the 2D case than that in the 3D case. This is perhaps unsurprising since in 2D the flow has fewer degrees of freedom to develop and evolve, resulting in violent fluctuations about the equilibrium state.

\begin{figure}
  \centerline{\includegraphics[width=\textwidth]{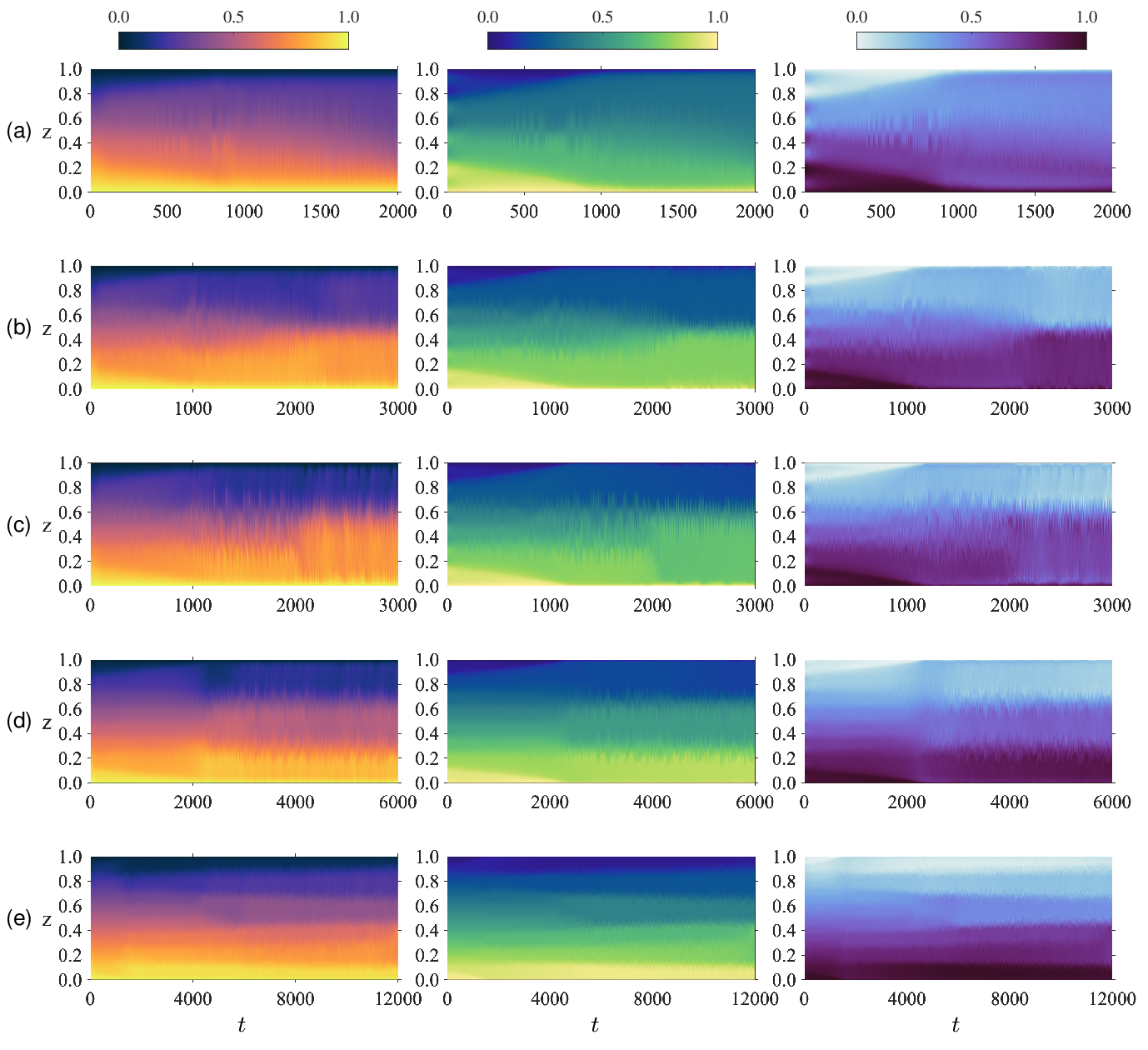}}
  \caption{Time evolution of horizontally averaged scalar profiles. (a-e) are for cases 1-3, 5, and 6, with parameters given in table \ref{tab:num}. Columns from left to right show mean profiles of non-dimensional temperature, salinity and density, respectively. Different cases share the same color map for each scalar field.}
\label{fig:profs}
\end{figure}

In figure~\ref{fig:profs} we plot the time evolution of the horizontally-averaged mean profiles of non-dimensional temperature $\langle\theta\rangle_h$, salinity $\langle s \rangle_h$ and density. Here $\langle\cdot\rangle_h$ stands for the spatial average over horizontal planes. With the non-dimensionalisation described in the previous section, temperature and salinity also both vary from one at the lower boundary $z=0$ to zero at the upper boundary $z=1$. Clearly, a robust staircase appears and survives for all five cases, but the time evolution varies significantly. Furthermore,  the strong oscillation of interfaces is apparent. For case 1 with the smallest $\Ray$, once the instability triggers the flow, a single layer quickly develops and extends over the entire interior of the flow. For cases with higher $\Ray$ the layering process is more complex. For case 2, starting from $t\approx1000$, two relatively weak interfaces appear in the center of the channel. The two interfaces rapidly merge into a single one at around $t\approx 2100$.

By comparing row 2 and 3 in the figure, it is again apparent that 2D and 3D simulations have essentially similar layering processes. As $Ra_\theta$ increases, more layers appear in the interior, as seen from the lower two rows of figure~\ref{fig:profs} for $Ra_\theta=10^7$ and $10^8$. For case 6, the simulation with the highest $Ra_\theta$ in our study, more layer merging occurs at later times. We continued the simulation until $t=12000$ and only four interfaces remained. It is important to note that the depth of the final layers is much larger than the characteristic depth of the initial perturbation $1/k$. Thus, we believe it is unlikely that the depth of the final layers is set by the vertical wavenumber of the initial perturbation. Rather, the depth of the final layers is due to the competition between different layers, as well as the confinement of the two boundaries.

\section{Vertical transport in layering state}\label{sec:transp}

Having considered qualitative aspects of the formation and survival of layered states, we now focus on the dynamics of case 5. This case clearly had a robust `staircase' at later times, and we are interested in the consequences of this staircase for the global transport and flow velocities. Throughout the simulation we record the appropriately scaled global heat and salt transfer rates as well as the flow velocity. These are quantified by appropriately defined Nusselt numbers and Reynolds numbers, 
\begin{equation}\label{eq:Nuss}
  \Nus_s = \frac{\langle u_z s \rangle_a}{\kappa_s \Delta_s H^{-1}},
 \quad\quad
  \Nus_\theta = \frac{\langle u_z \theta \rangle_a}{\kappa_\theta \Delta_\theta H^{-1}},
 \quad\quad 
  \Rey_i = \frac{u^{rms}_{i} H}{\nu}.
\end{equation}
Hereafter $\langle\cdot\rangle_a$ denotes the volume average over the whole flow domain. These volume averages in general depend on time.

\begin{figure}
  \centerline{\includegraphics[width=0.8\textwidth]{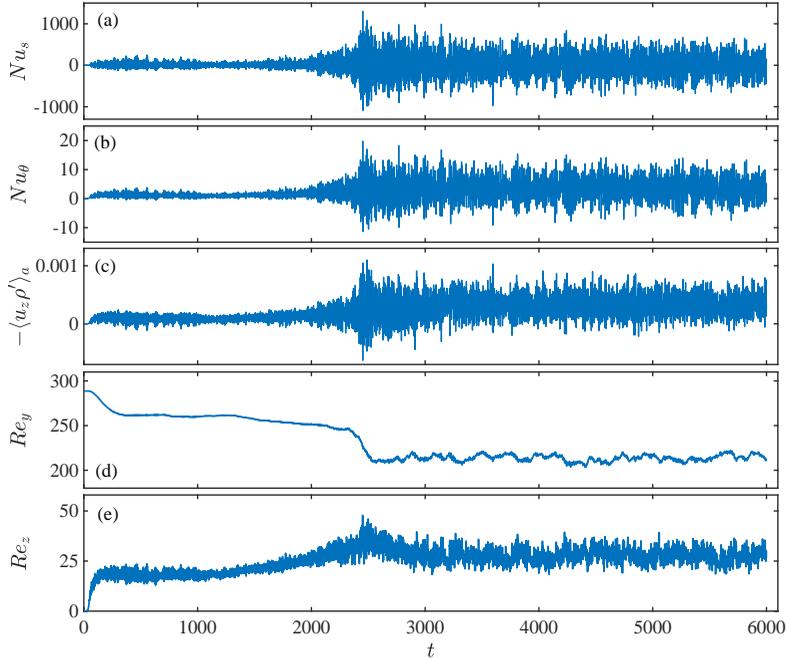}}
  \caption{Time history of (a) the scaled global salt transfer rate $\Nus_S$, (b) the scaled global heat transfer rate $\Nus_T$, (c) the convective density flux $-\langle u_z\rho' \rangle_a$, and (d, e) the Reynolds numbers $\Rey_y$ and $\Rey_z$, respectively, for case 5.}
\label{fig:tmhis}
\end{figure}

The three $u^{rms}_i$ are the root-mean-square values of the different components of the total velocity, including in particular the mean imposed shear. The density $\rho'$ and its associated volume-averaged convective flux $\langle u_z\rho' \rangle_a$ are also calculated. Figure~\ref{fig:tmhis} plots the time history of the global fluxes and Reynolds numbers for case 5 with $Ra_\theta=10^7$.

In Figure~\ref{fig:tmhis}, two qualitatively different stages  can be identified after the rapid initial spin-up of the flow, which  ends around $t=100$. The first stage occurs for  $t \lesssim 2000$. During this initial stage, the flow interior is dominated by KH-like shear-driven vortices. This stage ends when the flow relatively quickly reorganises into the layering state for $2000 \lesssim t \lesssim 2500$. During this transition period the KH-like vortices lose their spatially organized pattern and quickly break into turbulent convection layers. Such process can also be observed in the supplementary movies 5 and 6. Subsequently, global fluxes, as quantified by the Nusselt numbers, are both typically much larger in magnitude, and also fluctuate with a much larger magnitude. These fluctuations in the global quantities are caused by the strong oscillations of the interfaces, which can also be observed in the time history of the scalar mean profiles, as shown in figure~\ref{fig:profs}. The strong oscillation of the interfaces is induced by the violent convection motions in the adjacent layers. During the final stage, the equilibrium layering state can survive these strong oscillations for a very long time period. Also for our choice of control parameters the streamwise Reynolds number $\Rey_y$ is always significantly larger than the vertical Reynolds number $\Rey_z$, due to the contribution from the mean shear.

\begin{figure}
  \centerline{\includegraphics[width=\textwidth]{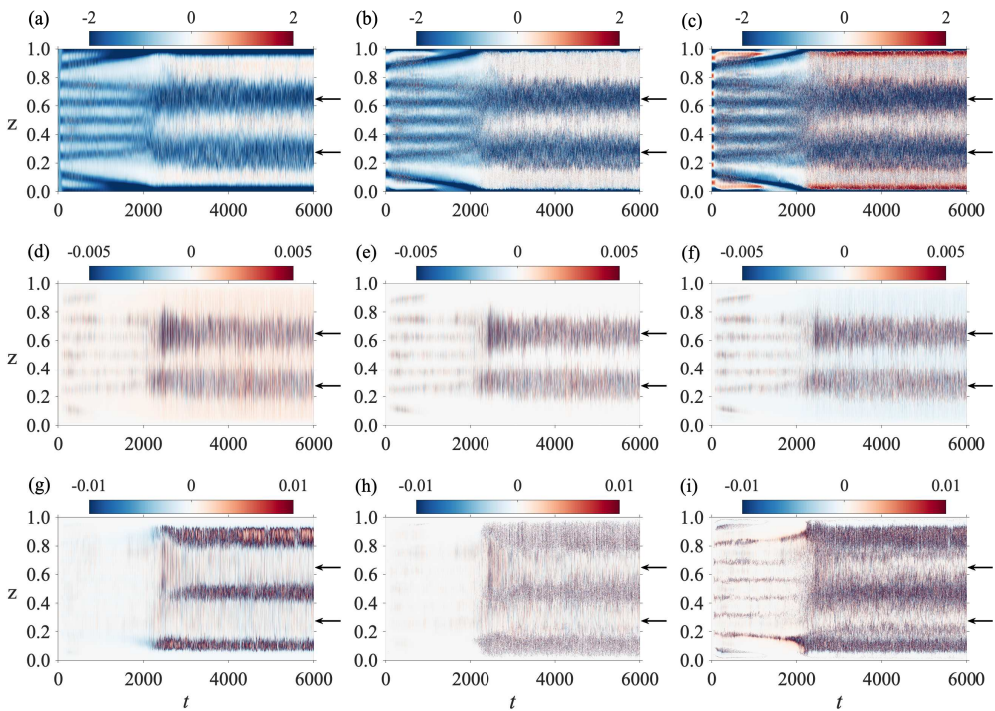}}
  \caption{Time evolutions of the horizontally averaged profiles for case 5. Rows from top to bottom show mean profiles of scalar gradients $\langle \partial_z\zeta \rangle_h$, scalar convective fluxes $\langle u_z\zeta \rangle_h$ and turbulent diffusivity $\kappa^T_\zeta = \langle u_z\zeta \rangle_h / \langle \partial_z\zeta \rangle_h$. Columns from left to right: $\zeta=\theta$, $s$, and $\rho'$, respectively. Black arrows mark the time-averaged heights of identifiable interfaces.}
\label{fig:flux}
\end{figure}

To reveal the effects of different structures on the transport properties, we calculated the time history of horizontally-averaged vertical profiles of scalar gradient $\langle \partial_z \zeta \rangle_h$, scalar convective flux $\langle u_z \zeta \rangle_h$ and turbulent diffusivity $\kappa^T_\zeta = \langle u_z\zeta \rangle_h / \langle \partial_z\zeta \rangle_h$, for the three scalars of interest with  $\zeta=\theta$, $s$, and $\rho'$. Figure~\ref{fig:flux} displays these time histories, once again for case 5. The staircase structure of thin interfaces and relatively deep layers is clearly apparent in the panels in the top row. The dark blue stripes mark the interfaces which have relatively high gradients of temperature, salinity, and density. The negative vertical gradient arises since both temperature and salinity decrease as the height increases, with salt stabilising and heat destabilising the flow statically. The relatively vertically extended white regions between interfaces with relatively weak gradients are the well-mixed layers with almost homogeneous mean temperature and salinity. The upper right panel of figure~\ref{fig:flux} indicates that inside the layers the density often exhibits a slightly positive net mean gradient, and it is this convectively unstable stratification which drives the large roll convection inside these layers, even though, overall, the density gradient is stable across the flow. 

As expected, the convective fluxes $\langle u_z\zeta \rangle_h$ for the three different scalars show very different behaviours in the interface and layer regions. The interface regions consist of strong oscillations between extreme positive and negative convective fluxes of heat, salinity, and density. This is essentially a geometric effect,  caused by the vertical spatial oscillations of the interfaces, in turn associated with the coupled effects of impinging vortices and ejecting plumes, as can be seen in the supplementary video. Nevertheless,  the interfaces with strong scalar gradients survive these spatial oscillations in our simulations. In the interior of  layers, oscillations are also observed in the convective fluxes, but overall $\langle u_z\theta \rangle_h$ and $\langle u_z s \rangle_h$ are positive, implying that both heat and salinity are transferred upward. Meanwhile, the convective flux of $\rho'$ is negative inside layers, indicating that the net density flux is actually downward in the layers as expected. This is consistent with the regular occurrence of static instability in the layers, and also with the density flux being dominated by the thermal convective component. This confirms that the shear has not  qualitatively affected the distinguishing characteristic of diffusive DDC of \emph{downwards} density flux. In the lower row of figure~\ref{fig:flux} we plot the mean profiles of turbulent diffusivities $\kappa^T_\zeta = \langle u_z\zeta \rangle_h / \langle \partial_z\zeta \rangle_h$. The most intense turbulent diffusivity actually occurs inside the layers, where  the magnitude of $\kappa^T_{\rho'}$ can be larger than $0.01$. It can also be noted that at the interface regions the turbulent diffusivities computed from horizontally averaged profiles have comparable values for temperature and salinity. However, within the layers the magnitude of $\kappa^T_\theta$ is much larger than $\kappa^T_s$. This implies that within the layers the flow is not in a fully developed turbulent state in our simulations, since for fully developed turbulent flow the convective fluxes should be similar for the two components.

\begin{figure}
  \centerline{\includegraphics[width=\textwidth]{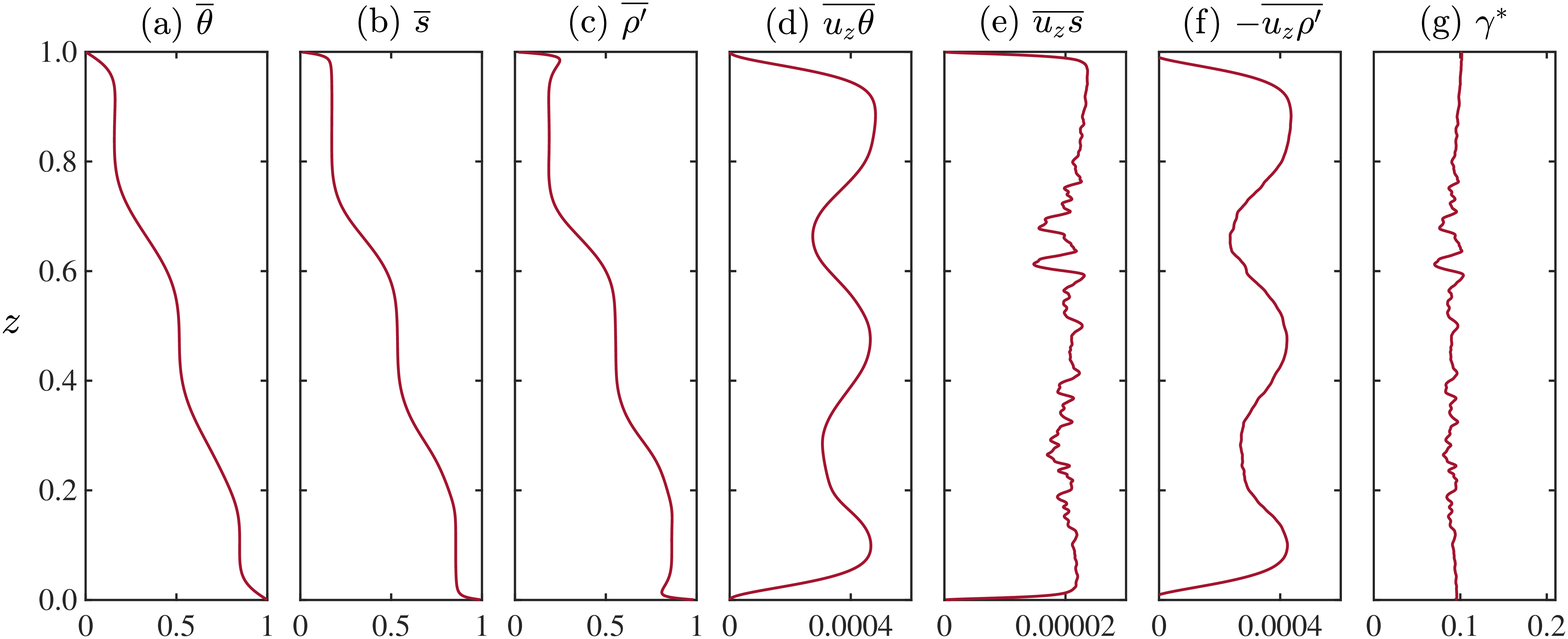}}
  \caption{Time averaged profiles for (a-c) the three scalar fields, (d-f) their convective fluxes, and (g) the total flux ratio  $\gamma^*$, as defined in (\ref{eq:gammastardef}). Time averaging is calculated from the data presented in figure~\ref{fig:flux} over the time interval $3000 \leq t \leq 6000$.}
\label{fig:fluxaver}
\end{figure}

In figure \ref{fig:fluxaver}, we plot time averages over the time period $3000<t<6000$ of the horizontally-averaged mean profiles shown in figure~\ref{fig:flux}, as well as the  flux ratio $\gamma^*$, defined as the ratio of the \emph{total} salt flux to the total heat flux:
\begin{equation}
  \gamma^*(z) = \frac{ \beta_s \left (\langle  w_{\rm dim} s_{\rm dim}  \rangle _h + \kappa_s \partial _z
      \langle s \rangle_h \right )}
  { \beta_\theta \left (\langle  w_{\rm dim} \theta_{\rm dim}  \rangle _h + \kappa_\theta \partial _z
      \langle \theta_{\rm dim} \rangle_h \right )} ,\label{eq:gammastardef}
\end{equation}
where the subscripts ${\rm dim}$ make explicit that the definition contains dimensional quantities and the total flux contains both convective and diffusive parts. Although in principle $\gamma^*(z)$ can vary with $z$, at a statistically stationary state both denominator and numerator should be constant with height.

Due to the vigorous and relatively high amplitude vertical oscillations of the interfaces, the time-averaged scalar profiles do not exhibit sharp interfaces but rather regions with relatively high gradient, near $z \simeq 0.3$ and $z \simeq 0.65$ as shown in the three leftmost panels of figure~\ref{fig:fluxaver}. Interestingly, although such spatial oscillations generate large, and yet crucially temporally-varying, local vertical fluxes as shown in figure~\ref{fig:flux}, after time averaging the interface regions exhibit smaller net convective heat flux. Upon time-averaging, the positive and negative fluxes with large magnitude visible in figure~\ref{fig:flux} associated with the significant convective deformations of the interfaces exhibit significant cancellation. Therefore, the net convective transport across the interfaces is substantially reduced. Conversely, the transport through the relatively well-mixed layers does not exhibit such variability, and in particular is typically single-signed, and so the net convective transport through the layers is significantly larger. Of course, this variability is counterbalanced by  the stronger diffusive flux through the high gradient interfaces. It is also important to appreciate that instantaneously, the transport across an interface can be very much larger in magnitude and opposite in sign to the net transport over a sufficiently long time interval, strongly suggestive of the physical mechanisms that lead to staircase structures having reduced net vertical transport.

Furthermore, the convective density flux is dominated by the heat flux. The net salinity flux is both significantly smaller in magnitude and also much closer to uniform across the entire flow, with no noticeable remnant remaining of the staircase structure in the time-averaged convective flux. Since the salinity is higher at the bottom boundary and lower at the top, it is reasonable to expect for the convective part of the salinity flux to be positive, i.e. in the opposite direction to the total salinity gradient across the whole fluid layer. Indeed, this is observed albeit the very small magnitude. Nevertheless, it is important to appreciate that the overall net convective density flux is negative, and hence downwards, even though the flow is statically stable, and so the flux here might be appropriately described as counter-gradient, in the sense that the net background density gradient is also negative, as is of course to be expected for DDC in the diffusive regime considered here. Interestingly, the middle layer is, upon this long-time averaging, statically stable, while the upper and lower layers are statically unstable. This qualitative difference is not reflected in the density flux within the layers, which is dominated by the vertical flux of heat, and largely takes similar values in each of the three layers. Clearly, from the time evolution data shown in figure \ref{fig:flux}, each of the layers is frequently statically unstable, and thus associated with vigorous convective overturning, and so it is not appropriate to draw any inferences from the long-time-averaged static stability of the middle layer.

Finally, the total flux ratio $\gamma^*(z)$ shown in the last panel of figure~\ref{fig:fluxaver} hardly varies with $z$ over the flow domain, giving further confidence to the assertion that the flow is essentially in a statistically steady state. More interestingly, $\gamma^* \simeq 0.1 =\sqrt{\tau}$, remembering the definition of the diffusivity ratio (\ref{eq:taudef}).  This is entirely consistent with the physical arguments presented by  \cite{Linden1978} considering idealized two-layer systems (see also \cite{Worster2004}). The  key insight presented by  \cite{Linden1978} is that, within the interfacial `boundary layers', salt diffuses in the thermally convective elements which then break away and are mixed into the `bulk' layers. This fundamental picture does not appear to be disrupted in any significant way by the presence of mean shear, augmenting the vortical scouring of the interfaces. Indeed, our results appear to give strong support to the variational arguments of \cite{Stern1982} that $\gamma^*$ should be bounded below by $\sqrt{\tau}$. Furthermore, our choice of density ratio  $\Lambda=2$ is both in the range $\Lambda < \tau^{-1/2}$ where the original model of \cite{Linden1978} admits steady  state solutions, and also at the start of the constant regime with uniform flux ratio $\gamma^* \simeq 0.13$ observed experimentally by \cite{Turner1965}. As discussed in detail by \cite{Worster2004}, there are several complicating issues in experimental realisations of diffusive DDC, whereas numerically it is straightforward to maintain constant boundary conditions for the salinity and temperature. 

Although the flux ratio does not appear to show any effect of the shear-driven forcing, i.e. not varying significantly in the vertical direction, it is still natural to be interested in the spatio-temporal properties of the scalar transport, particularly in such a controlled sheared and overall statically stable flow. In single-scalar stratified shear flows, it is commonplace to consider the `efficiency' of the shear-driven transport, which is traditionally quantified by an appropriately-defined turbulent flux coefficient, $\Gamma$. As introduced by \cite{Osborn1980}, $\Gamma$ is the ratio of the vertical density flux to the turbulent kinetic energy dissipation rate. In a statistically stationary flow where transport terms in and out of the domain can be ignored, and the turbulent kinetic energy is essentially constant, this ratio may be expected to quantify the relative strength of the density flux and viscous dissipation. For statically stable single-component flows then density flux may be related to irreversible changes in the potential energy. As here we are particularly interested in the dynamics of staircases, which exhibit spatial variation in the vertical direction, and specifically the transport properties of layers and interfaces, it is natural to define a flux coefficient as a function of $z$ and $t$ as
\begin{equation}\label{eq:quants}
\Gamma(z,t)= \frac{g\langle u_z\rho' \rangle_h}{\langle\varepsilon\rangle_h},
\end{equation}
where $\varepsilon$ is the local, pointwise dissipation rate of the total kinetic energy.

It is important to appreciate that, defined in this way, $\Gamma(z,t)$ is sign-indefinite due to sign-indefinite horizontally-averaged density flux in its numerator. $\Gamma$ will only be positive when, on average, dense parcels of fluid are being lifted up, as might be expected in statically stable turbulent shear flows with single component in a statistically steady state, but of course does not happen typically in diffusive DDC, even when sheared as we have observed above. Indeed, in single-component statically stable stratified flows, there has been much research devoted to attempts to go beyond the classical empirical observation of \cite{Osborn1980} that $\Gamma \lesssim 0.2$ to identify generic parameterisations of $\Gamma$ in terms of other parameters describing properties of the fluid, larger scale flow properties, and/or turbulence (see, for example \cite{Caulfield2021} for a review). Two particular parameters of interest are appropriately defined gradient Richardson numbers $Ri_g(z,t)$ and buoyancy Reynolds number $\Rey_b(z,t)$ 
\begin{equation}\label{eq:quants1}
  \Ric_g (z,t)= \frac{N^2}{S^2}, \quad \Rey_b(z,t) = \frac{\langle\varepsilon\rangle_h}{\nu N^2},
  \quad N^2 =\partial_z\langle g\rho' \rangle_h, 
\end{equation}
where 
for consistency with the chosen definition of $\Gamma$,  the buoyancy frequency $N^2$ and the background shear $S=\partial_z\langle u_y \rangle_h$ are defined here in terms of horizontally averaged quantities. Therefore both $\Ric_g$ and $\Rey_b$ may in general vary with vertical location and time, and as $N^2$ is only positive when the horizontally-averaged flow is statically stable, it is important to remember that both  $\Ric_g$ and $\Rey_b$ can in principle be negative, quite possibly within the layers, though the interfaces should still exhibit positive values. It should also be mentioned that $\Rey_b$ has been used to identify different regimes of mixing, namely a molecular regime, a transition regime, and a fully energetic regime~\citep{Ivey2008}.

\begin{figure}
  \centerline{\includegraphics[width=\textwidth]{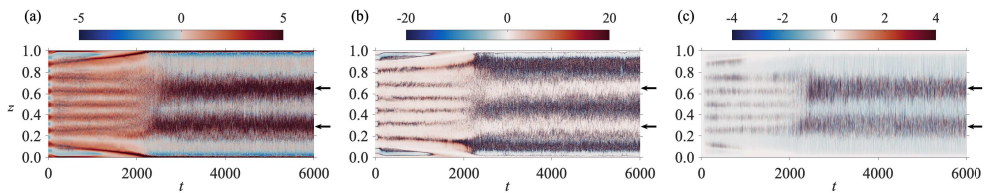}}
  \caption{Time evolutions for profiles of (a) $\Ric_g$, (b) $\Rey_b$, and (c) $\Gamma$, respectively, for case 5 with $Ra_\theta=10^7$. Black arrows mark the heights of interfaces. }
\label{fig:quants}
\end{figure}

Indeed, in figure~\ref{fig:quants} we show the time history for the profiles of the three quantities defined in (\ref{eq:quants}) and (\ref{eq:quants1}). As expected, after the initial transient to the asymptotic three-layer state, $\Ric_g$ and $\Rey_b$ display opposite behaviours in the interface and layer regions. Specifically, near interfaces $\Ric_g$ is large and $\Rey_b$ is small,  attributable to the high density gradients. Conversely, within layers $\Rey_b$ with large magnitude is associated with small in magnitude $\Ric_g$, with evidence of both positive and negative values for both parameters, corresponding to the convection motions. In the rightmost panel, it is evident that the ratio $\Gamma$ fluctuates very strongly near interfaces, attaining both positive and negative values with large magnitude. This suggests that the local dissipation rate being small near interfaces is dominating the observed numerical values, with the strong fluctuation of $\Gamma$ being clearly induced by the spatial oscillation of interfaces.
\begin{figure}
  \centerline{\includegraphics[width=0.7\textwidth]{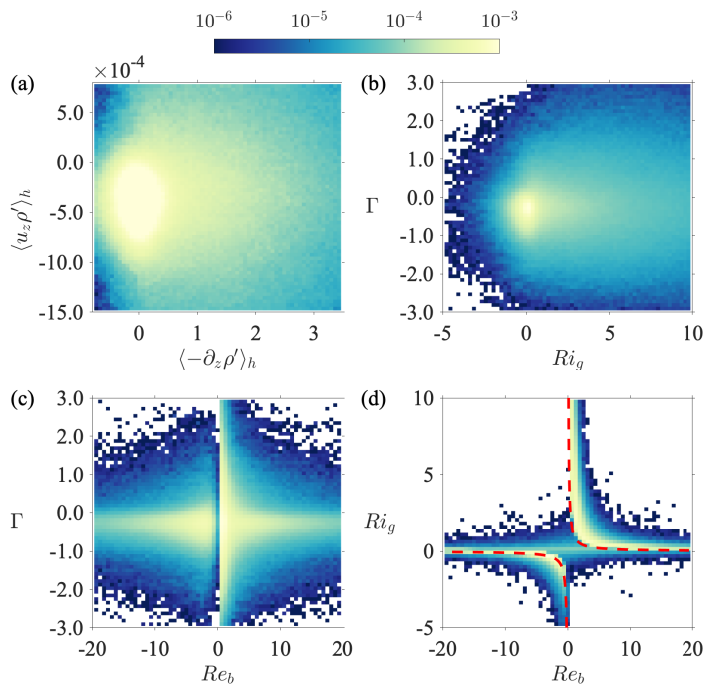}}%
  \caption{Joint relative probability for four pairs of flow quantities, including the vertical gradient of mean density $\langle -\partial_z \rho' \rangle_h$, the mean convective flux of density $\langle u_z\rho' \rangle_h$, $\Gamma=g\langle u_z\rho' \rangle_h / \rho_0\langle\varepsilon\rangle_h$, $\Ric_g=N^2/S^2$ and $\Rey_b = \langle\varepsilon\rangle_h/\nu N^2$. Statistics are calculated from the data of figure~\ref{fig:flux} for $t>3000$ and $0.1<z/H<0.9$. The color is shown on logarithmic scale.}
\label{fig:jpdf}
\end{figure}

A useful way to understand the relationship between the various quantities is to calculate the joint probability of pairs of quantities associated with specific values of $z$ and $t$. We present joint probability density functions (PDFs) for these data for four pairs of quantities in figure~\ref{fig:jpdf}. Figure~\ref{fig:jpdf}a shows the joint probability of $\langle -\partial_z \rho' \rangle_h$ and $\langle u_z\rho' \rangle_h$. Indicating that most of the flow domain is in relatively well-mixed layers, most of the data are associated with values of the mean density gradient being close to zero, though both positive and negative small gradients occur frequently. Conversely, there is a strong signal of predominantly negative density flux, associated with the expected strong convective flux of heat through the system, and particularly in the layers. Inside the layers the fluid is well mixed with nearly homogeneous mean scalars and so large scale convection rolls generate high convective flux. Unsurprisingly, there is asymmetry in the data associated with the existence of interfaces with high horizontally-averaged negative density gradient. The associated values of convective flux are spread over a wide range, which can be interpreted as being due to the significant spatial oscillation of interfaces.

In figures~\ref{fig:jpdf}b and c we show the joint PDFs of $\Gamma$ with $\Ric_g$ and $\Rey_b$, respectively. The $\Ric_g$ data in figure~\ref{fig:jpdf}b are once again dominated by values with relatively small magnitude, associated with most of the flow domain being in layers. Both negative and positive values are apparent, showing that the layers oscillated close to being well-mixed, while $\Gamma$ is clearly predominantly negative, since there is typically downwards convective density flux. Once again, there is asymmetry in the $\Ric_g$ data, with a wide range of positive values associated with the interfaces, associated also with a very wide range of different $\Gamma$ values.

Large magnitude values of $\Gamma$ are of course associated with significant density flux with relatively weak turbulence, precisely the behaviour which we would expected in the vicinity of a highly convoluted yet `sharp' interface dominated by scouring dynamics. Such apparently weak turbulence leading to significant convective transport is also apparent in  the $\Gamma(\Rey_b)$ data shown in figure~\ref{fig:jpdf}c. The data associated with the interfaces, with small positive $\Rey$ due to the combined effect of small $\varepsilon$ and large $N^2$ and a wide range of large magnitude $\Gamma$ is clearly apparent. Conversely, the layer data exhibit a wide range of both positive and negative $\Rey_b$ with predominantly smaller yet negative $\Gamma$.

Indeed, figures~\ref{fig:jpdf}b and c strongly suggest that $\Ric_g$ and $\Rey_b$ are anti-correlated, as can also be seen in figure~\ref{fig:jpdf}d. The expected asymmetry in data is clearly visible, where large positive $\Ric_g$ associated with interfaces is much more common than large negative values. The large positive $\Ric_g$ data correspond to small positive values of $\Rey_b$. Conversely, layers are associated with large in magnitude $\Rey_b$ data and small in magnitude $\Ric_g$, with both signs commonly occurring. We plot the heuristic relationship $\Ric_g \propto \Rey_b^{-1}$ with a red dashed line, which appears to be a reasonable approximation. This suggests that the dominant relationship is associated with the density distribution via the respective dependence on $N^2$ in the parameter definitions, and variations in the mean shear and dissipation rate only play a role at higher order. 

\begin{figure}
  \centerline{\includegraphics[width=0.5\textwidth]{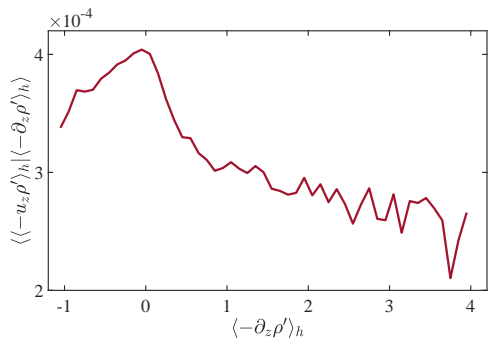}}
  \caption{Time average (over the intervale $3000 < t < 6000$) of the mean convection flux $\langle u_z\rho' \rangle_h$ conditioned on the mean density gradient $\langle -\partial_z \rho' \rangle_h$.}
\label{fig:cdavr}
\end{figure}

As a final demonstration of the properties of the convective fluxes of different flow regions, in figure~\ref{fig:cdavr} we plot the time-average over the interval $3000  < t < 6000$ of the negative of the density flux $\langle -u_z\rho' \rangle_h$ conditioned on $\langle -\partial_z \rho' \rangle_h$. A single peak appears with $\langle -\partial_z \rho' \rangle_h$ slightly smaller than zero, corresponding to regions of the flow with weak statically unstable density gradients. Such regions correspond naturally to the relatively well-mixed `layers'. As $\langle -\partial_z \rho' \rangle_h$ increases through positive values, the magnitude of the conditioned average broadly decreases. Higher values of the density gradient are naturally found in the interface regions, where the convective flux becomes relatively weak. As is apparent in figure~\ref{fig:flux}, the particular location of such  interfaces generically oscillates very strongly and it is therefore entirely reasonable that the  convective flux exhibits significant fluctuations. Nevertheless, the  conditioned average shown in  figure~\ref{fig:cdavr} suggests that the net overall convective density flux in the interface regions is  still negative, although the interfaces drive a downward density convective flux that is much weaker than that which is driven through the  layers. This is yet more evidence that the strong shear present in such flows does not modify the overall downward density flux characteristic of  diffusive DDC flows, due to the dominance of the upward convective heat flux.

\section{Development of local density inversion and multiple states of layering}\label{sec:multi}

\begin{figure}
  \centerline{\includegraphics[width=\textwidth]{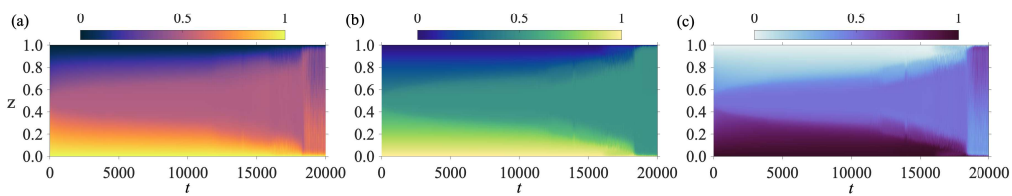}}
  \caption{Time evolutions of the horizontally averaged scalar profiles for case 7, starting from a single localized density inversion, with control parameters the same as case 5, as listed in table \ref{tab:num}. (a) Temperature, (b) salinity, and (c) density.}
\label{fig:newcsprofs}
\end{figure}

\begin{figure}
  \centerline{\includegraphics[width=\textwidth]{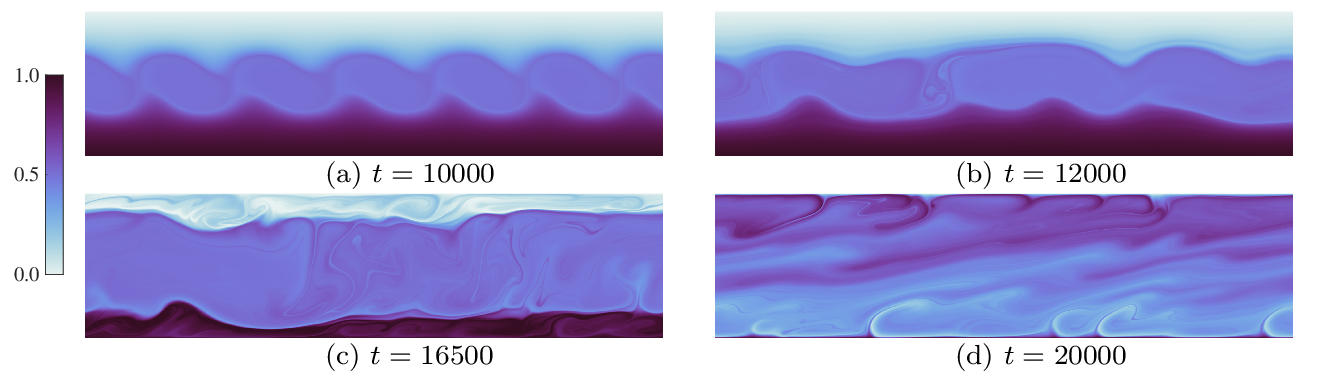}}
  \caption{Snapshots of density at four different time steps for case 7.}
\label{fig:newcsdens}
\end{figure}

Another interesting property of the sheared diffusive DDC flows considered here is that the final configuration of layering actually depends on the particular structure of the initial perturbation, in a non-trivial fashion. The co-existence of various layered states has been hypothesized by \cite{Stern1969} as metastable equilibria of the system. To demonstrate this, we consider the properties of case 7. As discussed in \S \ref{sec:method}, case 7 has the same flow control parameters as case 5 but a different, and more localised initial condition, given by (\ref{eq:case5n}) rather than (\ref{eq:initall}). Analogously to  figure~\ref{fig:profs},  we show the time evolution of the scalar profiles for case 7 in figure~\ref{fig:newcsprofs}. Also in figure~\ref{fig:newcsdens} we show the contours of density at four different times,. 

Initially, the single density inversion expands vertically over a sustained time period of more than 10000 non-dimensional time units. As is typical, Kelvin-Helmholtz like vortices grow within this expanding layer and then in turn break down into a well-mixed convection layer. Around $t=12000$, two new, essentially well-mixed layers emerge above and below the primary layer. However, these two relatively new layers are continually eroded, as is apparent in the bottom-left panel of figure \ref{fig:newcsdens}. Indeed, they do not survive and the main layer in the middle eventually occupies the whole channel, leading to a statistically steady state with a single convection layer, in qualitative contrast to the flow field shown in figure~\ref{fig:density} for the corresponding case 5. Therefore, for the same control parameters, different statistically steady states can be achieved starting from different initial density conditions. Without the presence of other layers, a single layer can grow continuously until the confinement of the boundaries sets in. Crucially, the various fluxes are also different for different states. The existence of multiple turbulent states with different transport properties has also been observed in other turbulent systems such as in Taylor-Couette flow~\citep{huisman2014} and fingering DDC staircases~\citep{yang2020}.

\begin{figure}
  \centerline{\includegraphics[width=0.9\textwidth]{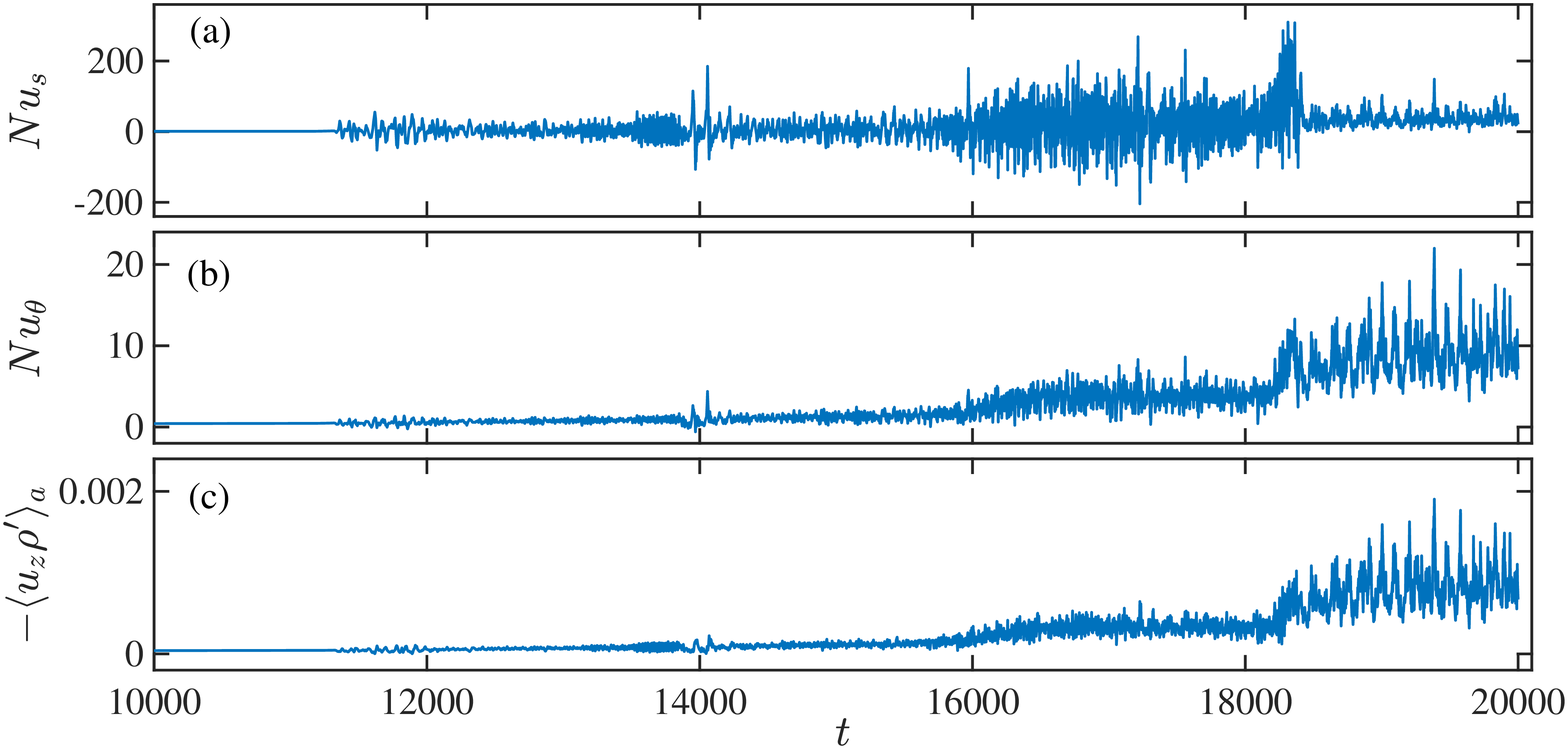}}
  \caption{Time history of (a) the scaled global salt transfer rate  $\Nus_S$, (b) the scaled global heat transfer rates $\Nus_T$, and (c) the convective density flux $-\langle u_z\rho' \rangle_a$ for case 7, as listed in table \ref{tab:num}.}
\label{fig:newcsflux}
\end{figure}

In figure~\ref{fig:newcsflux} we plot the time history of the two Nusselt numbers and the vertical flux of the density. During the multi-layer stage between $t=16000$ and $18000$, the fluxes fluctuate significantly. When the flow enters the single-layer final stage after $t>18000$, both the heat and density fluxes increase and fluctuate more strongly. The fluctuation of salinity flux, however, reduces significantly.

For cases 5 and 7 with $Ra=10^7$ and different initial perturbations, we measure the time-averaged fluxes at the final statistically steady stage for both the three-layer state of case 5 and the single-layer state of case 7. We average between $t=4000$ and $6000$ in figure~\ref{fig:tmhis} for the three-layer state, and between $t=19000$ and $20000$ in figure~\ref{fig:newcsflux} for the single-layer state. The two Nusselt numbers are $Nu_T=3.73$ and $Nu_S=20.9$ for the three-layer state, and $Nu_T=9.03$ and $Nu_S=37.1$ for the single-layer state, respectively. Therefore, under the same global control parameters, more layers and interfaces really significantly reduce the heat and salinity transport in the vertical direction, showing how important layering can be for `throttling' vertical fluxes, and so how crucial it is to understand whether a particular staircase arrangement is robust and long-lived in when subject to velocity shear, and hence further vigorous forcing. 

\section{Conclusions and outlook}\label{sec:conc}

In conclusion, we have conducted a series of simulations of sheared double diffusive convection (DDC) in the diffusive regime in a plane Couette flow geometry, using the Oberbeck-Boussinesq approximation so that the density depends linearly on both salinity and temperature.  The two bounding horizontal plates are maintained at constant temperature and salinity with the lower plate having higher temperature and salinity. Therefore the conductive salinity gradient is stabilising and the convective temperature gradient is destabilising, while the two plates have different velocities thus setting up a shear across the flow.  We choose the relative size of the temperature and salinity differences so that the flow is statically stable with density ratio $\Lambda=2$ and bulk Richardson number $\Ric_b=1$ as defined in (\ref{eq:lambda}) and (\ref{eq:ric}), respectively. We choose the Prandtl numbers to be $\Pra_\theta=10$ and $\Pra_s=1000$ for temperature and salinity respectively, so that the diffusivity ratio $\tau=100$, similar to oceanographic values.

For a variety of salinity and temperature Rayleigh numbers, and for both two- and three-dimensional simulations, we find that staircase structures spontaneously form, with relatively sharp interfaces of high scalar gradient separating relatively well-mixed and vigorously convecting layers. Generically, the formation of these layers appears to require a significant initial perturbation, in particular incurring static instability somewhere within the flow, and so, at least for this flow, it does not seem appropriate to consider the formation of these layers to arise from an infinitesimal linear instability of the flow system. Also, coarsening of the staircase structure after a significant period of highly disordered flow evolution is observed, further suggesting that the mechanisms by which the staircase structure loses (or alternatively retains) `stability' is highly nonlinear. Indeed, further evidence of the nonlinear complexity  of the structure of the solution space for this system arises from our demonstration that different initial (finite amplitude) perturbations can lead to different robust, attractive final staircase states.

Crucially however, we find that the injection of energy through shear does not modify certain key aspects of DDC flow in the diffusive regime. Staircase structures appear robust over very long time intervals. For the currently considered parameters, the enhanced mechanical forcing of the turbulence in the well-mixed layers due to the shear does not generically disrupt the staircase structure. This observation has implications for the robustness of double diffusive staircases in the presence of larger-scale shear, although the idealized nature of our flow geometry makes quantitative predictions of staircase survival challenging. Also, the power injection by the shear does not modify the domination of the density flux by the upward heat flux, so that the sheared diffusive DDC flows we have simulated exhibit downward vertical density flux on average.

On the other hand, our numerical simulations allow for the precise control of boundary conditions, a distinct advantage compared to laboratory experiments. With such control, we are able to obtain strong evidence in support of the mechanistic transport model proposed by \cite{Linden1978}, in that we show that the total flux ratio $\gamma^* \simeq \sqrt{t}$, as defined in (\ref{eq:gammastardef}). The physical picture of salt slowly diffusing into hot `blobs' of fluid that are then convected out of the boundary-layer-like interfaces is robust to the introduction of large-scale shear, which although it energises the turbulent flow does not seem to modify qualitatively or quantitatively the vertical scalar fluxes. The robustness of this balanced physical picture in particular suggests that for the control parameters chosen in this paper the dominant process for detachment of those hot `blobs' into the interior of the layers is dominated by convection, rather than shear-dominated vortical scouring.

A further attraction of numerical simulation is naturally the fact that the simulation data is in principle available at every point in space and time. We demonstrate the usefulness of this availability by considering horizontally-averaged data for various properties of the convective density flux, and in particular its relationship to two natural non-dimensional parameters to describe sheared stratified turbulence, namely the gradient Richardson number $\Ric_g(z,t)$ and the buoyancy Reynolds number $\Rey_b$ as defined in (\ref{eq:quants1}). Although the horizontal averaging inevitably smooths the properties of the interfaces which are significantly vertically perturbed, it is still possible to identify distinguishing characteristics of interfacial and layer regions. Specifically, the interfacial regions are distinguished by high positive local values of Richardson number, and low values of buoyancy Reynolds number, yet still non-trivial net downwards convective density flux. The net downwards density flux perhaps because of the smearing effect of horizontal averaging. Layer regions typically have larger downwards convective flux, and larger magnitude $\Rey_b$ with smaller magnitude $\Ric_g$. However, since they are close to well-mixed, substantial oscillations in sign of $\Ric_g$ are observed, without any particular correlation with the net downwards density flux. There is however clear anti correlation between $\Ric_g$ and $\Rey_b$, such that $\Ric_g \propto \Rey_b^{-1}$, suggesting that the mean velocity shear and the turbulent dissipation rate are not strongly affected by the prevailing stratification. This anti correlation is especially interesting when both quantities are positive, i.e. associated with regions of static stability.

Of course, these observations suggest several future avenues of investigation. Perhaps most pressing is to understand better the robustness of the various observed staircase flow states, and the evolution dynamics of the staircase layers. We have demonstrated that, depending on the initial structure of the perturbation, a flow with one set of global flow parameters can exhibit qualitatively different statistically stationary staircase states. The vertical flux properties of these different staircase states can be markedly different (with, as is well-known, the more layered state having substantially reduced convective transport). It is clearly of great practical interest to predict which possible staircase states are most likely to arise in any given situation, which is crucial for understanding the fate of diffusive staircases in the changing Arctic environments.

It is perhaps surprising that we have also shown that imposed mean shear with unity bulk Richardson number, which is still quite `strong' relative to the stable background stratification, still allows exceptionally long-lived staircase structures to survive. Therefore, it is a very interesting open question to identify criteria under which layered DDC flows in the diffusive regime can survive in the presence of externally imposed large scale flows, such as the mean shear flow considered here. Such criteria may be analogous to those of \cite{Taylor2017} for single-component stratification, and can be directly applied to oceanic situations. A first step in this direction would be to extend the present study to other Richardson numbers, which will the subject of a future work.

\backsection[Acknowledgements]{Y. Yang acknowledges the partial support by the Major Research Plan of National Natural Science Foundation of China for Turbulent Structures under the Grants 91852107 and 91752202. We also acknowledge the computing resources provided by PRACE on MareNostrum at Barcelona Supercomputing Center (BSC), Spain (Projects 2020235589 and 2020225335), and by the Gauss Centre for Supercomputing e.V. (www.gauss-centre.eu) on the GCS Supercomputer SuperMUC-NG at Leibniz Supercomputing Centre (www.lrz.de).}

\backsection[Declaration of interests]{The authors report no conflict of interest.}

\bibliographystyle{jfm}

\end{document}